# Single-shot polarization-resolved ultrafast mapping photography

Pengpeng Ding[1,#], Dalong Qi[1,#], Yunhua Yao[1], Yilin He[1], Jiali Yao[1], Chengzhi Jin[1], Zihan Guo[1], Lianzhong Deng[1], Zhenrong Sun[1], Shian Zhang,[1,2,3,*]

[1]*State Key Laboratory of Precision Spectroscopy, School of Physics and Electronic Science, East China Normal University, Shanghai 200241, China.*
[2]*Collaborative Innovation Center of Extreme Optics, Shanxi University, Taiyuan 030006, China.*
[3]*Collaborative Innovation Center of Light Manipulations and Applications, Shandong Normal University, Jinan 250358, China.*
[#]*The authors contributed equally to this work.*
[*]*Corresponding author. Email: sazhang@phy.ecnu.edu.cn*

**Abstract**

Single-shot ultrafast optical imaging plays a very important role in the detection of transient scenes, especially in capturing irreversible or stochastic dynamic scenes. To break the limit of time response speed of electronic devices, such as charge-coupled device (CCD) or complementary metal-oxide-semiconductor (CMOS) detectors, ultrafast optical imaging techniques usually convert the time information of a transient scene into the wavelength, angle, space or spatial frequency of the illumination light in previous studies. In this work, we propose a novel polarization-resolved ultrafast mapping photography (PUMP) technique by converting the time information into the polarization. Here, the spatiotemporal information of a dynamic scene is loaded into a rotationally polarized illumination laser pulse, and a polarization filtering in imaging detection and a deconvolution algorithm in image reconstruction are used to extract the original dynamic scene. In our PUMP system, the temporal resolution is 850 fs, the spatial resolution is 28.5 lp/mm at 700 μm × 700 μm field of view, and the number of frames is 16. By using PUMP, a spatiotemporal dynamics of femtosecond laser ablation in an indium tin oxide film on glass substrate is successfully captured. PUMP provides a new solution for measuring the transient scenes in a snapshot, which will bring a very wide range of applications in the field of ultrafast science.

**Keywords:** Ultrafast imaging; Image reconstruction; Polarization modulation; Laser ablation.

Ultrafast optical imaging (UOI) can capture the spatiotemporal information of dynamic scenes from nanosecond to femtosecond time scales, and therefore it has become a research hotspot in recent years [1-3]. So far, UOI has been successfully applied in various research fields, involving physics, chemistry, and biology [4-6]. Generally, there are two ways to achieve UOI, one is multiple-shot measurement based on a pump-probe strategy, and the other is single-shot measurement by the information conversion of the time to other dimensions. Multiple-shot UOI has been used to reveal various macroscopic and microscopic ultrafast dynamics [7, 8], but it is only applicable to the measurement of repeatable dynamic scenes, and it needs high stability in experimental parameters. Fortunately, single-shot UOI overcomes these limitations of multiple-shot UOI, which shows an irreplaceable role in detecting irreversible or stochastic dynamic scenes, such as plasma evolution [9], shock wave propagation [10, 11], and laser-matter interaction [12].

In the last decade, single-shot UOI has achieved leapfrog developments with the aid of optical field modulation or information multiplexing [1]. According to imaging type, single-shot UOI techniques can be further divided into active and passive imaging methods. For the single-shot active UOI, it requires a time-labelled illumination light to probe dynamic scenes, where various photon tags are utilized to resolve the time information of the dynamic scenes, including the wavelength, angle, space, and spatial frequency of the illumination light, such as sequentially timed all-optical mapping photography (STAMP) [13] and chirped spectral mapping ultrafast photography (CSMUP) [14] in the wavelength domain, Fourier-domain tomography (SS-FDT) [15] in the angle domain, femtosecond time-resolved optical polarimetry (FTOP) [16] in the space domain, and frequency recognition algorithm for multiple exposures imaging (FRAME) [17] in the spatial frequency domain. For the single-shot passive UOI, it needs an ultrafast planar-array detector to resolve the time information of dynamic scenes, such as ultrafast framing camera (UFC) [18] and compressed ultrafast photography (CUP) [19-22].

In above-mentioned single-shot active UOI techniques, the time information of the dynamic scene is mapped to the wavelength, angle, space or spatial frequency dimension of the illumination light. In this letter, we propose a novel polarization-resolved ultrafast mapping photography (PUMP) technique, where the time information is converted into the polarization dimension. A rotationally polarized laser pulse generated by an optical rotatory dispersion crystal (ORDC) is used as the illumination light of a dynamic scene, and a polarization filtering method and deconvolution reconstruction algorithm are jointly adopted to recover the original time and space information. In PUMP, the temporal resolution is 850 fs, the spatial resolution is 28.5 lp/mm at 700 μm × 700 μm field of view (FOV), and the number of frames is 16. To demonstrate the ultrafast imaging capability of PUMP, we successfully observe the spatiotemporal dynamics of femtosecond laser ablation in an indium tin oxide (ITO) film on glass substrate in experiment, and the observed dynamic behavior is highly consistent with previous theoretical model.

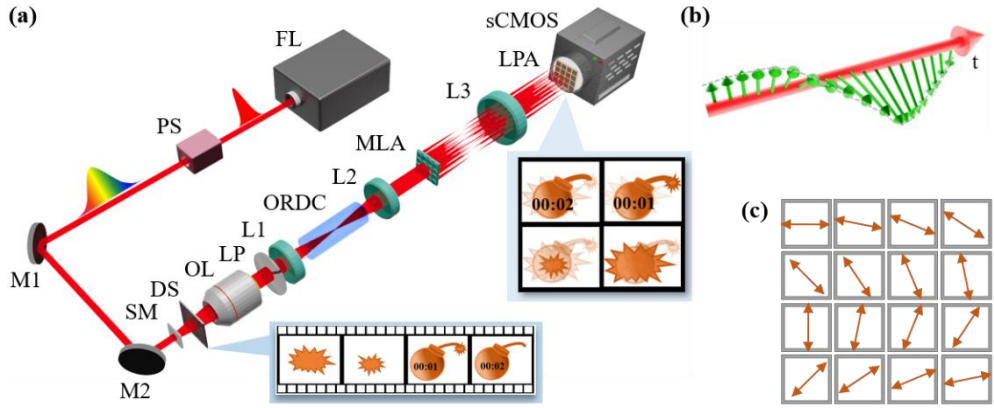

**Figure 1. Experimental arrangement of PUMP. (a)** Schematic diagram of PUMP. FL, femtosecond laser; PS, pulse stretcher; M, mirror; SM, scattering medium; DS, dynamic scene; OL, objective lens; LP, linear polarizer; ORDC, optical rotatory dispersion crystal; L, lens; MLA, micro-lens array; LPA, linear polarization array. The insets display the dynamic scene and integrated images measured by a sCMOS camera, respectively. **(b)** Polarization Axes of the illumination laser pulse after ORDC. **(c)** AoLPs of the sub-polarizers in LPA.

    The schematic diagram of PUMP is shown in Fig. 1(a). A Ti:sapphire femtosecond laser (FL) amplifier outputs the laser pulse with pulse duration of about 60 fs and central wavelength of 800 nm. The output laser pulse is firstly chirped in time by a pulse stretcher (PS), and then the chirped laser pulse transmits through a scattering medium (SM) and illuminates the dynamic scene (DS), thus the spatiotemporal information of the dynamic scene can be loaded into the chirped laser pulse. Here, the SM is used to reduce the spatial coherence of the chirped laser pulse, which can cover the aperture of each micro-lens in the array and achieve the sub-images of the dynamic scene with uniform FOV. Subsequently, the dynamic scene is imaged by an objective lens (OL) (Olympus, RMS10X-PF) with 10× magnification, and a linear polarizer (LP) (Thorlabs, WP25M-UB) is utilized to ensure that the chirped laser pulse is linearly polarized, and the angle of linear polarization (AoLP) can be controlled by rotating the polarizer. A customized ORDC of quartz (FuZhou MT-optics) with 140 mm length is placed in the center of a 4$f$ imaging system composed of a pair of lenses (L1 and L2) (Thorlabs, AC254-075-B) to modulate the polarization states of different frequency components in the chirped laser pulse, thus the time information from the low to high frequency components are encoded in different polarization states, and the formed rotational polarization in the chirped laser pulse is shown in Fig. 1(b). A 4 × 4 micro-lens array (MLA) (Highlight optics, ML-S1000-F60, pitch of 1 mm, focal length of 60 mm) is assembled on a customized 3D printed holder, which is placed on the focal plane (FP) of lens L2 to generate 16 sub-images. A field stop (not shown in Fig. 1(a)) is placed at the intermediate object plane of a 3.3× relay lens L3 (Thorlabs, MAP1030100-B) to avoid the crosstalk between adjacent sub-images. Close to the scientific complementary metal-oxide-semiconductor camera (sCMOS, Andor, Sona 4.2B) with 2048 × 2048 pixels, a customized 4 × 4 linear polarization array (LPA) is designed to match the beamlets generated by the MLA and acquire 16 time-encoded sub-images on the sCMOS camera. The 16 sub-polarizers in the LPA are obtained by cutting a linear

polarizing film (Edmund, 12-473) using a laser along different directions with an angle interval of 11.25°, and the AoLP arrangement of the LPA is shown in Fig. 1(c). The size of each sub-polarizer is 3 mm × 3 mm, and the 16 sub-images almost occupy the overall size of the sCMOS chip.

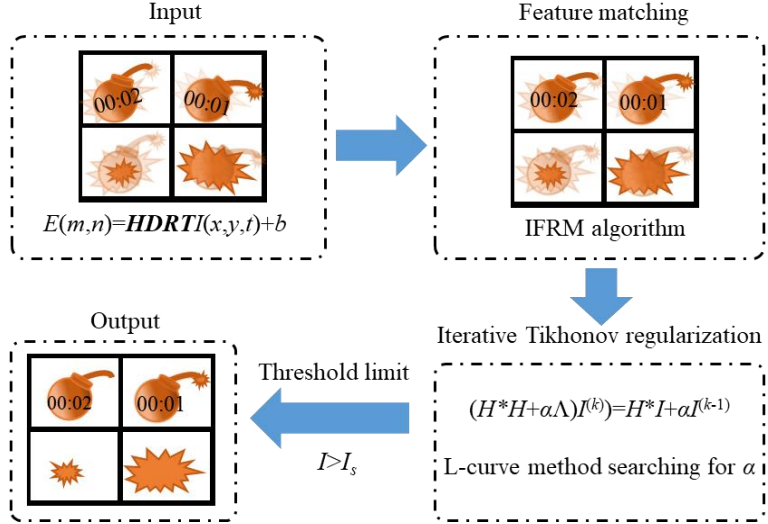

**Figure 2. Flowchart of the image reconstruction of PUMP**, involving feature matching, iterative optimization, and threshold limit.

The overall operation of PUMP is composed of image acquisition and reconstruction. In the image acquisition, the illumination laser pulse is firstly temporally chirped via a PS and rotationally polarized by an ORDC, then divided into multiple replicas by an MLA and filtered by an LPA, and finally measured by an sCMOS camera. It should be noted that, the illumination laser pulse carries the dynamic scene. In this process, the forward model can be described as

$$E(m,n) = \boldsymbol{HDRT}I(x,y,t) + b, \quad (1)$$

where $E(m, n)$ is the measured optical energy at pixel $m$, $n$ on the sCMOS camera, $I(x, y, t)$ is the original dynamic scene, and $b$ is the noise. Sequentially, $\boldsymbol{T}$ is the time-encoding operator determined by the time-polarization (TP) mapping relationship, $\boldsymbol{R}$ is the scene replication operator induced by the MLA, $\boldsymbol{D}$ is the optical distortion operator due to the differences in spatial positions among the $q$ lenslets in the MLA, and $\boldsymbol{H}$ is the circular convolution operator because of the polarization bandwidth of the LPA.

In the image reconstruction, an inverse problem of Eq. (1) involving deconvolution has to be solved. An iterative Tikhonov regularization [23, 24] is utilized to solve this problem by optimizing the following objective function, which can be expressed by

$$\arg\min_{I > I_S}\{\frac{1}{2}\|E - \boldsymbol{HDRT}I\|_2^2 + \alpha\Lambda(I)\}, \quad (2)$$

where $I_s$ is the threshold limit constant, $\|\cdot\|_2$ is the $l_2$ norm, $\alpha$ is the regularization parameter, and $\Lambda(I)$ denotes the Tikhonov regularization. The flowchart of the image reconstruction is shown in Fig. (2). The obtained two-dimensional image $E(m, n)$ is firstly segmented into multiple sub-images, and then an image feature recognition and matching (IFRM) algorithm is used to calibrate the optical distortion in each sub-image ascertained by a static test target in advance. Besides, to calibrate the intensity difference of these sub-images induced by the intensity distribution of the illumination pulse, optical properties of the object, and sensitivity of the CMOS, etc., each sub-image is pixel-wisely divided by a reference sub-image without an LPA. In the iterative optimization process after pre-processing, an L-curve method [25, 26] is used to search for the regularization parameter $\alpha$, and a convolution kernel is obtained by rotating the polarizer and measuring the transmitted intensity in experiment. After $k$ iterations of optimization, an intensity threshold limit is finally used to extract the estimated dynamic scene $\hat{I}(x, y, t_d)$, $t_d \in q$ [27].

Temporal and spatial resolutions are two important parameters in PUMP, which are jointly determined by the imaging hardware and reconstruction software here. Firstly, the temporal resolution is characterized, which can be obtained by measuring the time-wavelength (TW) and polarization-wavelength (PW) mapping relationships. Therefore, once the polarization resolution is obtained, the temporal resolution can be determined. In experiment, a chirped and scattered laser pulse with pulse duration of about 5 ps and central wavelength of 800 nm is spatially modulated by four hollow letters "**E**", "**C**", "**N**" and "**U**", and the AoLPs of the chirped laser pulse after the four letters are kept at 0°, 45°, 90° and 135° by inserting four polarizers, respectively. One sub-image of the spatially modulated laser spot with the four letters captured by PUMP without LPA is shown in the left pattern of Fig. 3(a). The four letters show uniform intensity distribution due to the lack of polarization modulation of the LPA. However, the intensities of the four letters are greatly modulated with the LPA, and the measured integrated intensities of the letters "**E**" and "**N**" in the 16 AoLPs of LPA are shown in the right pattern of Fig. 3(a), together with the reconstructed results obtained by formula (2). The two letters exist a difference of 90° in the peak intensity, which agrees with the difference of their polarization states. For the reconstructed curves after deconvolution, their full widths at half maximum (FWHMs) are greatly decreased compared with those of the measured results, and the averaged FWHM value is about 17°, which can be considered as the polarization resolution. The PW and TW mapping relationships are further explored, as shown in the left pattern of Fig. 3(b). To explore the PW mapping relationship, a polarizer with the same parameter in the LPA is fixed on a motorized rotating mirror holder (Thorlabs, PRM1Z8) with 0.1° angular resolution, and a spectrometer (Ocean optics, Maya 2000 pro) with 0.44 nm spectral resolution is utilized to measure the transmission intensities of the illumination laser pulse. A PW mapping curve is obtained, here the AoLP in each wavelength is calculated by $\psi = \tan^{-1}[\pm(2I_{45}/I_0)^{1/2} - 1]$, where $I_0$ is the intensity when the polarization axis of the polarizer is parallel to the wave vibration, i.e., $I_0 = I(\cos\psi)^2$, $I_{45}$ is the intensity when the angle turns to 45°, i.e., $I_{45} = I/2(\cos\psi + \sin\psi)^2$, and $I$ is the intensity without the polarizer. It is obvious that the PW mapping relationship keeps a good linearity, which does not affect the temporal resolution. For the TW mapping relationship, a customized frequency-resolved optical gating (FROG) system is used to achieve the TW mapping curve. One can see that the shorter and longer wavelength components, corresponding to the front and tail of the laser pulse, have the smaller dispersion coefficients than the central ones. Thus, the temporal resolution is mainly affected by the nonlinearity of the TW mapping relationship. Based on the reconstructed polarization resolution of 17°, the temporal frame interval in different wavelength components can be determined, as shown in the right pattern of Fig. 3(b). The temporal frame interval is much shorter in the front and tail of the laser pulse than that in-between, which is due to the nonlinear chirp in the time domain. Here, the temporal resolution, defined as the largest temporal frame interval, can be determined to be about 850 fs.

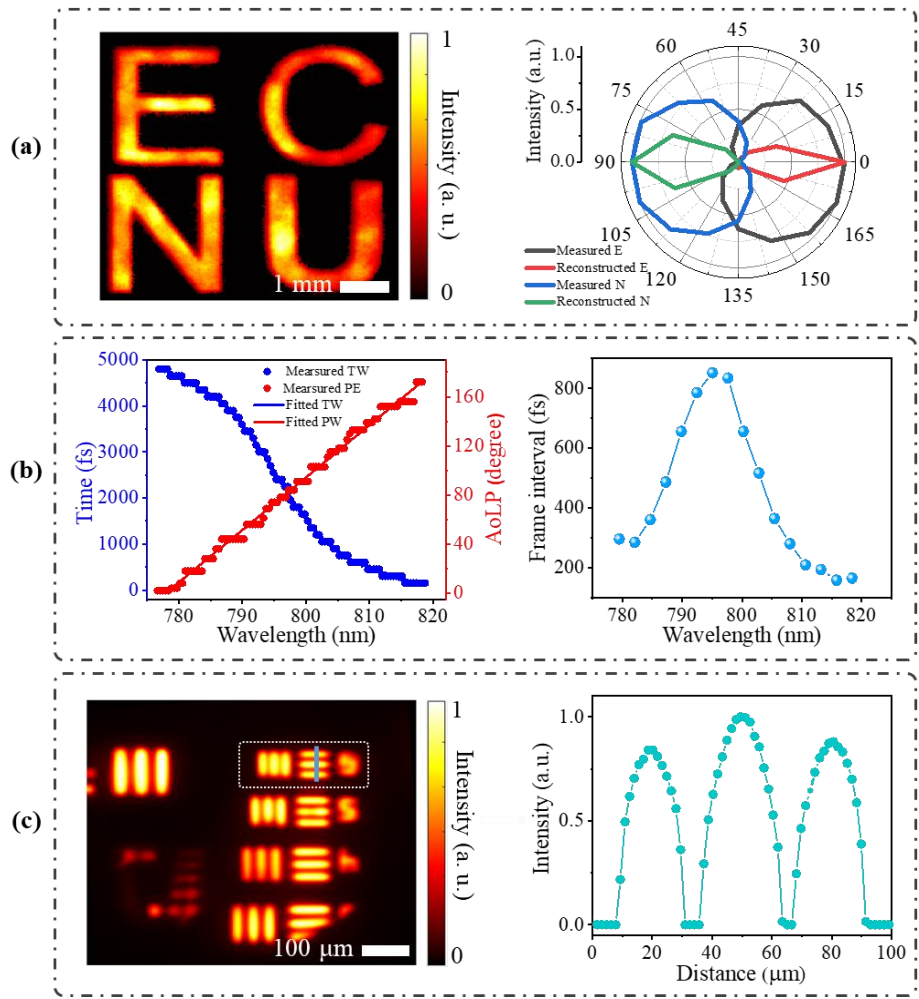

**Figure 3. Characterization of the temporal and spatial resolutions of PUMP. (a)** Detected image of letters "**E**", "**C**", "**N**" and "**U**" by PUMP without LPA (left) and measured and reconstructed integrated intensities of letters "**E**" and "**N**" with LPA (right). **(b)** Measured PW and TW mapping relationships for 5 ps illumination laser pulse (left) and calculated temporal resolution according to PW and TW mapping relationships (right). **(c)** Imaging of 1951 USAF resolution target for group 4 (left) and extracted intensity distribution of group 4 element 6 along the labelled line (right).

Secondly, the spatial resolution is characterized. In experiment, a USAF 1951 test target (Thorlabs, R3L3S1N) is placed on the object plane, and the measured sub-image is shown in the left pattern of Fig. 3(c). Obviously, the group 4 element 6 (labelled with a dashed box) can still be distinguished in both the horizontal and vertical directions. To display the spatial resolution more clearly, the intensities of the horizontal elements along the marked line are also calculated, and the obtained result is shown in the right pattern of Fig. 3(c). As can be seen, the three peaks are completely distinguished, which means that the spatial resolution is about 28.5 lp/mm at the FOV of 700 μm × 700 μm. By employing an objective lens with a higher numerical aperture or the lens L1 with a longer focal length, a higher spatial resolution can be obtained.

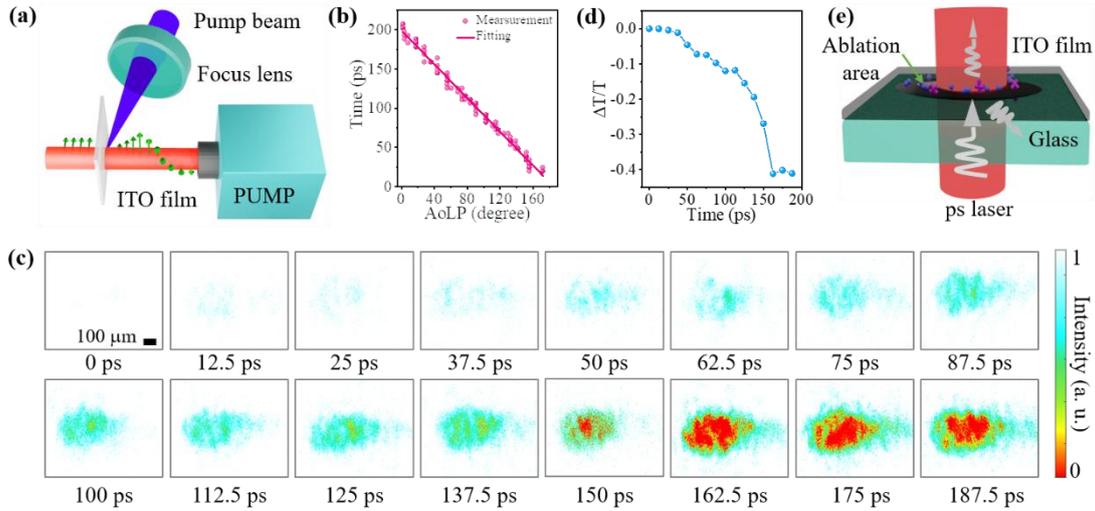

**Figure 4. Imaging of femtosecond laser ablation in ITO film by PUMP. (a)** Experimental design. **(b)** TP mapping relationship of 200 ps illumination laser pulse. **(c)** Spatiotemporal evolution of ITO film ablation. **(d)** Extracted transmittance difference ΔT/T of the ablation area. **(e)** Mechanism of femtosecond laser ablation in ITO film.

To demonstrate the ultrafast imaging capability of PUMP, the femtosecond laser ablation dynamics of an ITO film on glass substrate is experimentally explored. The thicknesses of the ITO film and glass are about 170 nm and 1 mm, respectively. As shown in Fig. 4(a), a 400 nm, 60 fs pump laser pulse is focused by a lens with 50 mm focal length, and then glanced at the ITO film with about 15° tilting angle to the sample surface. Simultaneously, a 200 ps illumination laser pulse is utilized to probe the ultrafast ablation scene induced by the pump laser pulse. It is worth noting that the ablation threshold of ITO [28] is more than one order of magnitude lower than that of glass [29], thus a suitable laser fluence is chosen to only excite the ITO sample. The fluence of the pump laser is about 0.23 J/cm$^2$, which does not reach the material removal threshold of ITO sample with 1.34 J/cm$^2$. The TP mapping relationship of the illumination laser pulse is shown in Fig. 4(b). As can be seen, the TP mapping curve shows a high linearity, which indicates that the temporal frame interval is uniform with 12.5 ps. The transmittivity dynamics process of the femtosecond laser-induced ablation in the ITO film captured by PUMP is shown in Fig. 4(c). The transmittance reduces within 162.5 ps and then almost keeps unchanged. To quantitatively analyze the dynamic behavior, a normalized integral transmittance difference ΔT/T in the ablation area of Fig. 4(c) is calculated, and the result is shown in Fig. 4(d). The value of ΔT/T shows a slow decrease followed by a fast decrease process, and finally it reaches the minimal value of -0.41 after 162.5 ps. A schematic diagram is given to explain the mechanism of femtosecond laser ablation in the ITO film, as shown in Fig. 4(e). Comprehensively, the physical mechanism of this ablation can be described as follows. In the beginning, the photon energy of the pump laser is absorbed to generate thermalized electrons, and then these thermalized electrons impart their energy into the ITO lattice structure, here the whole transfer duration is affected by the time of electron-phonon coupling. Immediately, a significant increase in free electron density leads to an increase of the reflectance for the illumination laser pulse, i.e., a decrease of the transmittance. Meanwhile, the continuous coupling of the electrons and phonons forms a liquid-gas mixture accompanied by absorption and scattering of photons on the film surface, which also causes a decrease of the transmittance [30]. After about 162.5 ps, the ablation area will gradually recover in hundreds of nanoseconds, which cannot be observed in this experiment due to the restriction of the time window. This recovery of the ablation area is verified by a second measurement without the pump laser.

Considering the unique imaging strategy, PUMP is equipped with the following five advantages: i) The polarization encoding method of PUMP completes the missing piece left by existing ultrafast imaging methods using spectral and spatial dimensions; ii) The ability of flexible detection by precisely controlling the AoLPs of the sub-polarizers in PUMP enables it to accurately capture the spatial information at a specific instant within the dynamic scene; iii) Different from STAMP and CSMUP, PUMP does not require the large spectral bandwidth, because the range of the AoLP can be flexibly adjusted by changing the length of the ORDC; iv) Benefiting from the wide spectral response of the ORDC (200-2500 nm for quartz), PUMP can well work for illumination lasers with various wavelength ranges, especially for the ultraviolet and infrared lasers; v) PUMP can also be switched into spectral imaging mode based on the PW mapping relationship as shown in Ref. [31], which can greatly extend its applications.

To conclude, we have developed a single-shot PUMP technique by converting time information into polarization. In PUMP, the spatiotemporal information of a dynamic scene is loaded into a rotationally polarized illumination laser pulse, and then it is extracted by a polarization filtering in imaging detection and a deconvolution algorithm in image reconstruction. The temporal resolution, spatial resolution, and number of frames in the PUMP system are 850 fs, 28.5 lp/mm at 700 μm × 700 μm FOV and 16, respectively. Moreover, the spatiotemporal dynamics of femtosecond laser ablation in an ITO film is successfully captured by PUMP to demonstrate its powerful capability in ultrafast detection. This novel technique provides a well-established tool for measuring the transient scenes in a snapshot, which will promote the advances in both fundamental and applied sciences [32-34].

**Acknowledgements**

This work was partially supported by the National Natural Science Foundation of China (91850202, 92150301, 12074121, 62105101, 62175066, 12274129, 12274139), and the Science and Technology Commission of Shanghai Municipality (21XD1400900, 21JM0010700, 20ZR1417100).

# Supplemental material for "Single-shot polarization-resolved ultrafast mapping photography"

Pengpeng Ding[1,#], Dalong Qi[1,#], Yunhua Yao[1], Yilin He[1], Jiali Yao[1], Chengzhi Jin[1], Zihan Guo[1], Lianzhong Deng[1], Zhenrong Sun[1], Shian Zhang,[1,2,3,*]

[1]*State Key Laboratory of Precision Spectroscopy, School of Physics and Electronic Science, East China Normal University, Shanghai 200241, China.*
[2]*Collaborative Innovation Center of Extreme Optics, Shanxi University, Taiyuan 030006, China.*
[3]*Collaborative Innovation Center of Light Manipulations and Applications, Shandong Normal University, Jinan 250358, China.*
[#] *The authors contributed equally to this work.*
[*]*Corresponding author. Email: sazhang@phy.ecnu.edu.cn*

# Details of the image reconstruction for PUMP

**i) Forward imaging model.**

The overall forward imaging model of PUMP can be described as

$$E(m,n) = \boldsymbol{HDRT}I(x,y,t) + b, \tag{1}$$

where $E(m, n)$ is the measured optical energy at pixel $m, n$ on the sCMOS camera, $I(x, y, t)$ is the original dynamic scene, and $b$ is the noise. Sequentially, $\boldsymbol{T}$ is the time-encoding operator determined by the time-polarization (TP) mapping relationship, $\boldsymbol{R}$ is the scene replication operator induced by the MLA, $\boldsymbol{D}$ is the optical distortion operator due to the differences in spatial positions among the $q$ lenslets in the MLA, and $\boldsymbol{H}$ is the circular convolution operator because of the polarization bandwidth of the LPA. To show the mathematics in detail, we use $I(x, y, t_d)$ to show the detected frames:

$$I(x, y, t_d) = \int K_d(\psi) \times \boldsymbol{DRT}I(x,y,t)d\psi, \tag{2}$$

where $t_d$ is the sequence number, $K_d(\psi)$ is the convolution kernel and each element of $K_d(\psi)$ is the intensity ratio when photons transmit through a polarizer with specific AoLP.

**ii) Backward image reconstruction.**

In PUMP, the spatiotemporal information of each frame is also contributed by the other 15 frames due to the overlap in the polarization, so a deconvolution algorithm is utilized to retrieve the exact information of each frame. Thus, an inverse problem of Eq. (1) involving deconvolution has to be solved. Here, an iterative Tikhonov regularization is utilized to solve this problem by optimizing the following objective function, which can be expressed by

$$\arg\min_{I>I_S}\{\frac{1}{2}\|E - \boldsymbol{HDRT}I\|_2^2 + \alpha\Lambda(I)\}, \tag{3}$$

where $I_s$ is the threshold limit constant, $\|\bullet\|_2$ is the $l_2$ norm, $\alpha$ is the regularization parameter, and $\Lambda(I)$ denotes the Tikhonov regularization. With the objective function established, the image reconstruction can be divided into the following five steps:

**Step 1. Image segmentation.** The 2D image $E(m, n)$ with 2048 × 2048 pixels detected by an sCMOS camera is firstly segmented into 16 frames with 512 × 512 pixels.

**Step 2. Calibration of the optical aberration.** To eliminate the optical aberration, a static test target is measured by PUMP in advance, and then an image feature recognition and matching (IFRM) algorithm built in MATLB computer vision toolbox is used to quantitatively determine the distortion degrees between the reference frame in the center and the other 15 frames. Once these distortion degrees are quantitatively determined through the transformation matrixes between the reference frame and the other 15 frames, these matrixes can be used to match the spatial locations with pixel-by-pixel correspondence.

**Step 3. Calibration of the intensity difference.** To eliminate the intensity difference induced by the intensity distribution of the illumination light, optical properties of the object, sensitivity of CMOS, etc., each frame was pixel-wisely divided by a reference frame without an LPA, and an intensity calibration matrix is obtained to extract the real intensities of the 16 frames.

**Step 4. Iterative deconvoluting optimization.** The optical energy at each pixel from the 16 frames with corresponding spatial location, i.e., $b_{i,j} = I(x_i, y_j, t_d)$, $t_d \in [1, 16]$, $i, j \in [1, 512]$, is chosen for iterative deconvolution. For convenience, the deconvolution of the operator $\mathbf{H}$ can be expressed by

$$Ax_\alpha = b_{i,j}, \qquad (4)$$

where $x_\alpha$ is the solution, $A$ is the circular convolution kernel matrix with each row obtained by cyclically shifting the convolution kernel. The condition number of $A$ is very large, which cannot be dealt using an inverse matrix, and therefore an iterative Tikhonov regularization is used, here the L-curve method is employed to search for the regularization parameter $\alpha$. By introducing the discrepancy of $d_\alpha = Ax_\alpha - b_{i,j}$ and the weighting parameter of $\varphi$ obeying from a monotonically increasing function, the L-curve can be defined as

$$L = \left\{ (\varphi(\|x_\alpha\|^2), \varphi(\|d_\alpha\|^2)) : \alpha > 0 \right\}. \qquad (5)$$

Thus, the regularization parameter $\alpha$ is corresponding to the value at the corner of the L-curve. After $k$ iterations, the optimized deconvolution result is obtained.

**Step 5. Threshold limitation.** To further reduce the convolution effect and improve the polarization resolution, a hard-threshold constant is used to limit the reconstructed $x_\alpha$, which is an empirical parameter depending on the actual experimental results. In this operation process, the 16 reconstructed data points are firstly normalized, then the data points below the threshold are eliminated, and finally the modulated normalized data points are multiplied with the corresponding original data points to retrieve the dynamic scene.